\documentclass[conference]{IEEEtran}
\IEEEoverridecommandlockouts

\usepackage{cite}
\usepackage{amsmath,amssymb,amsfonts}
\usepackage{graphicx}
\usepackage{textcomp}
\usepackage[T1]{fontenc}
\usepackage[narrow,varqu,var0]{zi4}
\usepackage[protrusion=true,expansion=true]{microtype}
\usepackage{xcolor}
\usepackage{listings}
\usepackage{booktabs}
\usepackage{needspace}
\usepackage[hyphens]{url}
\usepackage[hypertexnames=false,breaklinks=true,hidelinks]{hyperref}
\usepackage{xurl}
\usepackage{tikz}
\usetikzlibrary{arrows.meta, positioning, shapes.geometric, patterns}

\hypersetup{
  pdftitle={Ground-Side Mission Plan Compilation with Policy-as-Code Guardrails for Cloud-Native Satellite Platforms},
  pdfauthor={Hsiu-Chi Tsai, Chia-Tung Chung},
  pdfsubject={IEEE SMC-IT/SCC 2026},
  pdfkeywords={satellite mission planning, OPA/Rego, Argo Workflows, Kueue, ORCHIDE}
}

\usepackage{float}

\def\BibTeX{{\rm B\kern-.05em{\sc i\kern-.025em b}\kern-.08em
    T\kern-.1667em\lower.7ex\hbox{E}\kern-.125emX}}

\begin{document}

\title{Ground-Side Mission Plan Compilation with Policy-as-Code Guardrails for Cloud-Native Satellite Platforms}

\author{
\IEEEauthorblockN{Hsiu-Chi Tsai}
\IEEEauthorblockA{National Yang Ming Chiao Tung University\\
Hsinchu, Taiwan\\
thc1006@ieee.org}
\and
\IEEEauthorblockN{Chia-Tung Chung}
\IEEEauthorblockA{National Yang Ming Chiao Tung University\\
Hsinchu, Taiwan\\
jun.514114.ee10@nycu.edu.tw}
\thanks{\textcopyright~2026 IEEE. Personal use of this material is permitted. Permission from IEEE must be obtained for all other uses, in any current or future media, including reprinting/republishing this material for advertising or promotional purposes, creating new collective works, for resale or redistribution to servers or lists, or reuse of any copyrighted component of this work in other works. Accepted for publication in the IEEE Space Mission Challenges for Information Technology / Space Computing Conference (SMC-IT/SCC), 2026. This arXiv version is an extended version of that paper, adding a unified GPU+CPU DRA quota and a scheduler-level accelerator fallback re-validated on Kueue~v0.18.3 (Section~\ref{sec:dra-unified}).}
}

\maketitle

\begin{abstract}
Onboard cloud-native runtimes for satellites are emerging on multiple tracks (ORCHIDE's K3s and Argo Workflows stack, Axiom Space's AxDCU-1 with Red Hat MicroShift, and Kepler Communications' NVIDIA Jetson Orin edge nodes), but each assumes that the workflow artifacts it executes arrive from the ground. ORCHIDE's architecture document D3.1 states explicitly that ``only the Deferred Phase is part of the ORCHIDE scope,'' and no open-source ground-side toolchain has been released by the consortium. We present \textsc{Satellite Mission Compiler}, a four-stage pipeline that addresses this gap: it takes a human-authored mission plan, checks it against machine-checkable structural and policy rules, and compiles it into the container-workflow artifacts that cloud-native satellite runtimes consume. Concretely, the pipeline parses the plan against a Pydantic schema derived from public ORCHIDE materials, evaluates it against an OPA/Rego policy package of ten deny rules with explicit ORCHIDE-inherited versus author-imposed provenance, compiles it into a typed WorkflowIntent intermediate representation (IR), and renders it as Argo Workflow directed acyclic graphs (DAGs) and Kueue Job manifests with Dynamic Resource Allocation (DRA) support. We classify pre-uplink loss events into four severity tiers tied to specific schema and policy checks, and anchor the layered-validation design in the safety reading of defense-in-depth (NASA-STD-8739.8B), distinguished from the security reading of NIST~SP~800-53. The implementation is validated by golden translation evaluations, \texttt{argo lint}, an in-process baseline that reproduces OPA's decisions, and live single-node cluster submission---including a DRA-backed GPU admission cascade on Kueue~v0.17.3 (re-validated on v0.18.3) and, on v0.18.3, a unified GPU+CPU device-class quota with a scheduler-level accelerator fallback. Six Model Context Protocol (MCP) tools expose the pipeline to AI agents through a path-traversal-protected interface (CWE-22). The compiler is released under EUPL-1.2 (DOI~10.5281/zenodo.21228150).
\end{abstract}


\section{Introduction}
\label{sec:introduction}

Onboard cloud-native runtimes for satellites have moved from concept to deployment along several parallel tracks. The ORCHIDE EU project (Horizon grant~\#101135595) has publicly specified an onboard stack of \texttt{K3s}, Argo Workflows, \texttt{containerd}, \texttt{urunc}~\cite{urunc}-hosted unikernels, EOS storage, and \texttt{vAccel}-mediated accelerators running on heterogeneous onboard hardware (NVIDIA Jetson Orin GPUs, Xilinx Versal FPGAs, NXP LX2160 ARM64 CPUs, and Kalray MPPA accelerators), presented at KubeCon~EU~2026~\cite{orchide-d31,orchide-kubecon,edhpc2025}. Red~Hat and Axiom~Space's AxDCU-1, an orbital data-center prototype powered by Red~Hat Device Edge (which includes the MicroShift lightweight Kubernetes distribution) was announced in March~2025~\cite{microshift-axdcu1} and delivered to the International Space Station aboard SpaceX~CRS-33 in August~2025 for an on-orbit data-processing demonstration. Kepler Communications reported in March~2026 the commissioning of distributed on-orbit compute across its ten Tranche~1 satellites, using forty NVIDIA Jetson~Orin modules as edge-compute nodes interconnected by optical inter-satellite links to support AI and accelerated workloads~\cite{kepler-jetson}. The technology stacks differ in important ways, and KubeSpace~\cite{kubespace} has argued that the standard Kubernetes control plane is ill-suited to low-Earth-orbit (LEO) satellite networks characterized by geographic dispersion and frequent handovers. What the deployed tracks share, however, is the assumption that the workflow artifacts they execute---directed acyclic graphs (DAGs), container manifests, resource claims---arrive from the ground.

The artifacts arrive, but the ground-side tooling that should produce them is uneven. ORCHIDE's architecture document D3.1 states explicitly that ``only the Deferred Phase is part of the ORCHIDE scope'' (Section~\ref{sec:orchide-background}): the onboard Orchestrator executes mission plans but does not itself generate, validate, or compile them. The consortium's ground segment does include a Software Development Kit (SDK), a Simulation Framework, and a Management Framework, but none of those tools, nor any mission-plan schema or policy-enforcement layer, has been released as open source (Section~\ref{sec:orchide-background}). The other onboard tracks similarly assume but do not provide a way to take a structured mission plan, prove it is well-formed under a published schema, and emit the Argo artifacts the onboard runtime executes together with the Kueue artifacts that gate ground-side admission. The consequence of operating without that layer is concrete rather than hypothetical: a plan whose acquisition step has no fallback resource class declared can reach the satellite and stall the Deferred Phase when the requested accelerator is unavailable; a download event missing a ground-visibility window can be admitted and then never executed, queuing the next plan behind it; a structural error in the YAML payload can require a corrective uplink that consumes the next contact window. Section~\ref{sec:loss-events} classifies these consequences by severity tier.

Validation of mission plans against configurable rules is not a new problem. Commercial mission-planning systems such as GMV FlexPlan~\cite{flexplan} have for decades supported reusable constraint and rule definitions, and command-sequence checking against flight rules is a mature practice at NASA's Jet Propulsion Laboratory (JPL), exemplified by the RP-check architecture used in Mars-rover operations~\cite{rpcheck}. Open-source frameworks exist on the ground-control side (OpenC3~COSMOS~\cite{openc3} runs on Kubernetes via Helm, scaling to constellation-sized fleets) and the EOEPCA platform~\cite{eoepca} already compiles Common Workflow Language specifications for Earth-observation data-processing applications into Argo workflows. Each of these systems addresses part of the problem the ORCHIDE-class onboard runtimes raise; none of them, however, provides as an open-source toolchain the specific combination of (i)~policy-as-code over (ii)~satellite \emph{mission plans}, distinct from generic command sequences or data-processing apps, (iii)~compiled into Argo Workflow and Kueue Job artifacts that (iv)~target the artifact formats consumed by onboard cloud-native runtimes such as ORCHIDE.

This paper contributes that synthesis. We present \textsc{Satellite Mission Compiler}, an open-source four-stage pipeline that takes a structured mission plan and produces validated Argo and Kueue artifacts. Concretely:
\begin{itemize}
  \item A \texttt{Pydantic}~\cite{pydantic} schema derived from the public ORCHIDE materials enumerated in Section~\ref{sec:orchide-background}, with parse-time enforcement of timezone-aware timestamps, instrument presence for acquisition events, and structural integrity of AI services and steps.
  \item An OPA (Open Policy Agent)/Rego policy package of ten deny rules, separated explicitly into rules derived from ORCHIDE materials and rules imposed by the authors as safety practice, with provenance indicated in Table~\ref{tab:rules}.
  \item Argo Workflow and Kueue Job renderers supporting sequential and parallel execution modes, soft-preference GPU affinity, and Dynamic Resource Allocation (DRA) for accelerators. DRA-backed GPU admission with single-device quota was demonstrated against Kueue~v0.17.3 on a kubeadm cluster (Section~\ref{sec:dra-quota}), and extended on Kueue~v0.18.3 to a unified GPU+CPU device-class quota and an opt-in scheduler-level \texttt{firstAvailable} accelerator fallback, with the empirical boundary that Kueue quota-counts only \texttt{exactly} device requests (Section~\ref{sec:dra-unified}).
  \item Six Model Context Protocol (MCP)~\cite{mcp-spec} tools that expose the pipeline to AI agents through a path-traversal-protected interface (Common Weakness Enumeration~CWE-22).
\end{itemize}

The implementation is validated by golden translation evaluations, static Argo linting against \texttt{argo}~v4.0.1, an in-process baseline that reproduces the OPA policy decisions, and live-cluster submission on host kubeadm Kubernetes clusters (v1.35--v1.36 across experiments) running Argo, Kueue, and the NVIDIA DRA driver. Source code is released under the European Union Public Licence v1.2 (EUPL-1.2) at \url{https://github.com/thc1006/satellite-mission-compiler} (DOI~\href{https://doi.org/10.5281/zenodo.21228150}{10.5281/zenodo.21228150}).

\section{Background and Related Work}

\subsection{Background: The ORCHIDE Onboard Cloud-Native Platform}
\label{sec:orchide-background}

ORCHIDE (Orchestration of Reliable Computing on Heterogeneous Infrastructures Deployed at the Edge) is an EU Horizon~Europe project running from December~2023 to May~2026, led by Thales~Alenia~Space with Tarides, KP~Labs, Thales~Romania, and POLITEHNICA Bucharest as partners~\cite{orchide-d31}. Its objective is an onboard Edge Platform-as-a-Service for Earth-observation satellites: a runtime stack that lets operators deploy, update, and revise containerized AI workflows while the spacecraft is in orbit. The publicly described ORCHIDE solution consists of three components targeted at the ground segment---a Software Development Kit, a Simulation Framework, and a Management Framework---together with the onboard runtime they target, the Orchestrator~\cite{orchide-d22,orchide-d31}. The compiler presented in this paper provides three functions absent from the published ORCHIDE artifacts: schema validation, policy-as-code enforcement, and compilation into the artifacts the Orchestrator consumes.

\paragraph*{Onboard technology stack.} ORCHIDE's onboard Orchestrator runs \texttt{K3s}~\cite{k3s} as its lightweight Kubernetes distribution and Argo Workflows~\cite{argo} as the directed-acyclic-graph execution engine, with \texttt{containerd} and the \texttt{urunc}~\cite{urunc} OCI (Open Container Initiative) runtime hosting \texttt{Unikraft} and \texttt{MirageOS} unikernels under hypervisor isolation. The platform uses EOS, the distributed-storage system originally developed at CERN, as its data backbone; Zot as the OCI registry; and \texttt{vAccel} as a hardware-agnostic abstraction over graphics-processing-unit (GPU) and field-programmable-gate-array (FPGA) accelerators~\cite{edhpc2025,orchide-eos-cern}. The choice is constrained by the size, weight, and power (SWaP) envelope typical of an Earth-observation payload; the use of Argo Workflows specifically allows a mission plan to be represented as a graph of containerized steps with explicit data and resource dependencies.

\paragraph*{Structure of an ORCHIDE mission plan.} A mission plan is a time-ordered sequence of events distributed over one or more orbits. Each event has a type, either an \emph{acquisition}, which is bound to an instrument identifier (e.g., optical and hyperspectral cameras, radars, or multi-detector arrays) and a region descriptor distinguishing ocean (O) from land (L)~\cite{edhpc2025,orchide-kubecon}, or a \emph{download}, characterized by a ground-visibility window and a transmission duration. Acquisition events carry one or more \emph{AI services}; each service is annotated with a priority on the 1--4 scale used by the onboard scheduler~\cite{orchide-kubecon} and is decomposed into an ordered sequence of \emph{steps}. Steps declare a phase (\textit{preprocessing}, \textit{AI inference}, or \textit{postprocessing}) and a resource class (CPU, GPU, or FPGA)~\cite{orchide-kubecon}. Plans may also declare an optional fallback resource class per step. The public ORCHIDE materials note this field but do not specify how an Orchestrator ought to act on it; this paper treats fallback declaration as a recommended safety practice and enforces it through a policy rule for any accelerator-requesting step (Section~\ref{sec:policy}, Rule~4)---an author-imposed rule, not an ORCHIDE-inherited requirement.

\paragraph*{The Deferred Phase scope and the ground-side gap.} D3.1 states explicitly that ``only the Deferred Phase is part of the ORCHIDE scope'' and that the Acquisition and Transmission phases are outside that scope~\cite[\S 2.2]{orchide-d31}. The mission plan itself reaches the Orchestrator through D3.1's mission-plan deployment interface, identified as \texttt{IF\_SO\_MIS\_DP} in the project's interface catalogue~\cite[\S 3.3.1, Table]{orchide-d31}, and is consumed by the onboard Mission Manager and dispatched to its Workflow Manager subcomponent for execution~\cite[\S 3.2.1.1]{orchide-d31}. The Orchestrator does not generate, validate, or compile that plan; although the consortium has demonstrated graphical SDK tools on its YouTube channel~\cite{orchide-yt-wfb}, no open-source ground-side toolchain, mission-plan schema, or policy-enforcement layer has been released as of June 2026. A malformed plan that bypasses ground-side checks consumes scarce in-orbit compute and downlink budget before the next plan can be uploaded, with consequences ranging from missed observation windows to Deferred-Phase stalls. Section~\ref{sec:loss-events} returns to this consequence model. Section~\ref{sec:overview} presents the compiler that addresses this gap.

\paragraph*{What public materials cover and what they do not.} Our schema and compilation pipeline are derived from the union of publicly available ORCHIDE artifacts to date: deliverables D2.2 (state-of-the-art review of edge orchestration and unikernel technologies), D3.1 (overall solution architecture and design), and the dissemination deliverables D6.1 and D6.2~\cite{orchide-d22,orchide-d31,orchide-d61,orchide-d62}; Karam~Hankache's EDHPC~2025 talk on the onboard data-processing orchestrator~\cite{edhpc2025}; the KubeCon~EU~2026 talk by Karam~Hankache and Weisz~\cite{orchide-kubecon}; the CERN-hosted workshop presentation by Weisz on ORCHIDE's use of EOS for space edge computing~\cite{orchide-eos-cern}; and the consortium's June~2026 archival paper~\cite{weisz2026orchide}, published after our initial submission, which documents the onboard architecture but does not release a ground-side schema or policy package. The technical deliverables D3.2, D4, D5, and D7 remain confidential at the time of writing. The \texttt{Pydantic} schema we present in Section~\ref{sec:schema} is designed as a pluggable boundary: when an authoritative ORCHIDE specification or an open-source schema is released, replacement is intended to be mechanical because the schema does not propagate into the policy package or into the renderers. Section~\ref{sec:limitations} returns to this point.

\subsection{Policy-as-Code, Safety, and the Security--Safety Distinction}
\label{sec:policy-rationale}

OPA~\cite{opa} is a general-purpose policy engine that evaluates rules written in the Rego declarative language. OPA is a Cloud Native Computing Foundation (CNCF) Graduated project and is widely deployed as a Kubernetes admission controller through Gatekeeper~\cite{gatekeeper}. We use OPA because the policy is a separate artifact that a reviewer can execute and audit standalone: an external auditor can run \texttt{opa eval} against arbitrary inputs without invoking the compiler. Version control and independent review OPA shares with any in-repo validator; what OPA adds over a hand-written Python validator is compiler-independent executable evaluation delivered by a standard, off-the-shelf external tool.

\paragraph*{Layered defensive validation.} The compiler enforces four independent checks on every mission plan before any artifact is admitted to a cluster: (i)~structural validation against a \texttt{Pydantic}~v2 schema, (ii)~semantic policy evaluation against ten Rego deny rules, (iii)~static lint of the rendered Argo Workflow YAML via \texttt{argo lint}, and (iv)~runtime admission by the Kubernetes API server, optionally fronted by Kueue. We call this, by analogy with NASA-STD-8739.8B's layered Independent Verification and Validation (IV\&V), \emph{layered defensive validation by partially redundant checkers}: each stage has a distinct primary failure mode---the schema catches type, cardinality, and required-field errors at parse time; the policy catches cross-field semantic errors that span multiple fields; the static lint catches manifest-level errors introduced by the renderer; and admission catches violations of the target cluster's own constraints---and the stages overlap by design so that a defect in an overlapping check does not silently propagate. The schema and policy layers overlap on five of the ten rules (1, 2, 7, 8, and~9), documented in Section~\ref{sec:policy}; the resulting pre-uplink loss-event classification is given in Section~\ref{sec:loss-events}.

\paragraph*{Security defense-in-depth versus safety defense-in-depth.} The term \emph{defense-in-depth} is used in two distinct technical communities and carries different operational interpretations in each. In the security community, defense-in-depth refers to overlapping controls (network, host, application, and data) that together mitigate adversarial threats to confidentiality, integrity, and availability, as catalogued in NIST~SP~800-53~\cite{defense-in-depth}. In the safety community, defense-in-depth refers to layered, mutually redundant verification of safety-relevant artifacts so that no single analytical or implementation error can propagate undetected to the operational system. The safety reading is articulated by NASA's Software Assurance and Software Safety Standard~\cite{nasa-std-8739-8} and is consistent with the constraint-enforcement perspective of Leveson's Systems-Theoretic Accident Model and Processes (STAMP)~\cite{leveson-stamp}, whose associated hazard-analysis technique is Systems-Theoretic Process Analysis (STPA). The two readings are complementary but not interchangeable, and the layered validation pipeline of this paper instantiates the safety reading through partially redundant layers. The ORCHIDE consortium's own archival paper~\cite{weisz2026orchide} describes its onboard Security Manager as offering a ``layered defense-in-depth posture'' against an ``adversarial satellite environment''---verbatim the security reading---which the ground-side validation pipeline of this paper does not duplicate. The compiler does not implement the security reading---it does not authenticate uplink operators, encrypt transmitted artifacts, or defend the cluster against an adversarial mission-plan author---and that scope boundary is recorded in Section~\ref{sec:limitations}.

\paragraph*{Scope disclaimer.} NASA-STD-8739.8B~\cite{nasa-std-8739-8} prescribes a full software-assurance program including hazard analysis, assurance records, and IV\&V by an organization independent of the developer; this paper does not produce those deliverables. STPA-style hazard analysis would be the natural successor to the consequence-tier classification we adopt in Section~\ref{sec:loss-events} but is outside the scope of a ground-side compiler. We invoke both standards as the source of the safety-defense-in-depth practice that motivates the layered design, not as compliance claims.

\paragraph*{On novelty.} OPA satisfies the version-control, decoupling, and external-audit requirements that motivate our choice of policy engine. The contribution of this paper is the synthesis of OPA-style policy-as-code with a typed mission-plan schema and an Argo/Kueue renderer targeting the artifact formats of an onboard cloud-native runtime; the precise scope of that synthesis claim, and the adjacent systems it does and does not overlap with, are deferred to Section~\ref{sec:related-work}.

\subsection{Kubernetes Batch Admission}

Kueue~\cite{kueue} is a Kubernetes-native job queuing system providing fair-share scheduling, resource quotas, and preemption for batch workloads. Baseline DRA integration (the \texttt{Resource\allowbreak Claim\allowbreak Template} path with \texttt{device\allowbreak Class\allowbreak Mappings}) shipped as alpha in Kueue~v0.14.0 (September~2025); v0.17.0 (March~2026) added an alpha extended-resources path enabled by the \texttt{DRA\allowbreak Extended\allowbreak Resources} feature gate. The v0.17.3 release~\cite{kueue-v0173} (May~2026) that we exercise in Section~\ref{sec:dra-quota} retains the same gate~\cite{kueue-dra-extended-resources}. Volcano~\cite{volcano} provides a richer abstraction (scheduler, PodGroup, queue), but its API surface exceeds the requirements of single-Job-per-service workloads. We treat Kueue as the admission target because its ClusterQueue + LocalQueue abstraction maps cleanly to the per-mission quota model the renderer needs; the application to satellite workload scheduling is the contribution Section~\ref{sec:related-work} positions among adjacent ground-side toolchains.

\subsection{Related Work}
\label{sec:related-work}

\paragraph*{Onboard cloud-native runtimes for satellites.} Several tracks are converging on container-based orchestration in orbit, each naming a specific Kubernetes-derived or Kubernetes-adjacent runtime rather than vanilla Kubernetes. The ORCHIDE consortium specifies an onboard \texttt{K3s}+Argo stack (detailed in Section~\ref{sec:orchide-background}) with a project-end demonstration that was targeted for May~2026~\cite{orchide-d31,orchide-kubecon,edhpc2025}. Red~Hat and Axiom~Space's AxDCU-1, an orbital data-center prototype running Red~Hat Device~Edge (which includes the MicroShift lightweight Kubernetes distribution) was announced in March~2025~\cite{microshift-axdcu1} and delivered to the International Space Station aboard SpaceX~CRS-33 in August~2025 for on-orbit data-processing experiments. Kepler Communications reported in March~2026 the deployment of forty NVIDIA Jetson~Orin modules across its ten Tranche~1 satellites; as of that announcement the orchestration layer above the Jetson modules is not publicly specified~\cite{kepler-jetson}. KubeSpace's orbit-aware critique~\cite{kubespace} targets onboard control-plane stability and is orthogonal to the ground-side compilation layer the present work occupies.

\paragraph*{Commercial mission planning and command-sequence validation.} Validation of mission plans against configurable rules is a mature practice. GMV's FlexPlan is a commercial mission-planning product that has supported reusable constraint and rule definitions for ground-segment scheduling across multiple operators~\cite{flexplan}. On the spaceflight-operations side, JPL's RP-check is an architecture for command-sequence validation against flight rules, developed and deployed in Mars-rover surface operations~\cite{rpcheck}. Both reflect the practitioner expectation that automated rule checking is essential before commands or plans leave the ground.

\paragraph*{Ground-control and EO data-processing frameworks.} On the open-source side, OpenC3~COSMOS is an open-source mission-operations framework, with a separate commercial Enterprise edition that adds Kubernetes/Helm deployment for constellation-scale fleets; its focus across both tiers is telemetry, command, and the operator-facing console rather than mission-plan compilation~\cite{openc3}. The European Space Agency (ESA) EOEPCA platform compiles Common Workflow Language~(CWL) specifications for Earth-observation data-processing applications into Argo workflows via Calrissian~\cite{eoepca}, but the inputs are processing-application descriptors, not satellite mission plans, and policy enforcement is not part of the compilation contract. Adjacent systems address different layers: Krios~\cite{krios} and Komet~\cite{komet} contribute LEO scheduling and serverless abstractions, KubeEdge~\cite{kubeedge} provides edge orchestration with reported satellite deployments, Kratos~OpenSpace~\cite{kratos} deploys Rancher Kubernetes Engine~2 (RKE2)/\texttt{K3s} for ground-station signal processing, and the Consultative Committee for Space Data Systems' mission-planning standard (CCSDS-MPS)~\cite{ccsds-mps} together with operator systems such as the German Aerospace Center (DLR) EnMAP Mission Planning System~\cite{enmap-mps} formalise constraint-based mission planning at a layer above the workflow runtime.

\paragraph*{Recent developments (2026).} The literature published while this system matured advances the layers around the compiler without closing the ground-side gap. On the space cloud-native side, YUHENG-OS proposes a cloud-native operating system for distributed satellite clusters and benchmarks it against vanilla Kubernetes~\cite{yuhengos}, while Equinox turns onboard battery, thermal, and queue state into a marginal cost of execution for decentralized scheduling on Earth-observation constellations~\cite{equinox}; both sit at the onboard-runtime and scheduling layer that consumes workflow artifacts rather than the ground-side layer that produces and validates them. A position paper on the edge-cloud-space continuum enumerates the assumptions that break when serverless and orchestration models move to low-Earth orbit and names constraint-aware placement of workflow graphs as an open problem~\cite{edgespace2026}---the placement our compiler front-loads with schema and policy checks. On the heterogeneous-scheduling side, an independent Kubernetes+Argo+Kueue framework for hybrid quantum-classical pipelines corroborates the toolchain we chose for accelerator admission~\cite{hybridqc2026}, ClusterLess demonstrates deadline-aware serverless workflow orchestration across federated edge clusters~\cite{clusterless}, and BIDENT maps neural-network operators across heterogeneous CPU/GPU/NPU units on edge hardware~\cite{bident}---an operator-level analogue of the GPU$\rightarrow$CPU device fallback of Section~\ref{sec:dra-unified}. On the workload side, edge intelligence for satellite Earth observation schedules image acquisition and on-board-versus-ground processing under energy constraints~\cite{eosched2026}, the acquisition-and-download task model our mission plans encode. Finally, an empirical study of policy-as-code adoption across open-source projects frames OPA/Rego governance as an established engineering practice rather than an ad hoc choice~\cite{pacadoption}; space-specific policy-as-code nonetheless remains scarce, a gap this paper's policy package addresses. These systems collectively cover the onboard-runtime, edge-orchestration, heterogeneous-scheduling, and policy-as-code layers that surround the compiler, yet none takes a satellite mission plan through schema and policy validation into onboard-runtime artifacts.

\paragraph*{The ground-side compilation gap.} None of the four threads above offers a typed, policy-checked compilation path from a satellite mission plan to onboard-runtime workflow artifacts. The onboard-runtime thread stops at the runtime itself; the commercial-planning and command-sequence threads define rules but not container-workflow output; the ground-control and EO threads either focus on telemetry-and-command (COSMOS) or compile from processing-application descriptors rather than mission plans (EOEPCA); and the adjacent constraint-based planning systems (KubeSpace/Krios/Komet/CCSDS-MPS/EnMAP) operate at a layer above the workflow runtime without emitting validated Argo or Kueue artifacts. Section~\ref{sec:overview} presents the compiler that addresses this gap; Table~\ref{tab:comparison} quantifies the gap across the dimensions of mission-plan input, policy-as-code, and Kubernetes-artifact output.

\begin{table*}[htbp]
\caption{Comparison with Adjacent Systems}
\label{tab:comparison}
\centering
\small
\begin{tabular}{@{}lcccc@{}}
\toprule
System & Mission plan input & Policy-as-code (PaC) & K8s artifact output & Domain \\
\midrule
\textbf{This work} & \checkmark & \checkmark (OPA/Rego) & \checkmark (Argo + Kueue) & Satellite ground-side \\
ORCHIDE~\cite{orchide-kubecon} & Receives plan & -- & Argo (embedded) & Satellite onboard \\
AxDCU-1 / MicroShift~\cite{microshift-axdcu1} & -- & -- & -- & Orbital data-center prototype \\
Kepler / Jetson~\cite{kepler-jetson} & -- & -- & -- & On-orbit Jetson edge compute \\
KubeSpace~\cite{kubespace} & -- & -- & -- & LEO control plane \\
Krios~\cite{krios} & -- & -- & -- & LEO scheduling \\
Komet~\cite{komet} & -- & -- & FaaS (custom) & LEO serverless \\
GMV FlexPlan~\cite{flexplan} & \checkmark (commercial) & proprietary rules & -- & Ground-segment planning \\
JPL RP-check~\cite{rpcheck} & \checkmark (cmd. seq.) & flight rules (non-PaC) & -- & Command-sequence validation \\
OpenC3 COSMOS~\cite{openc3} & -- & -- & K8s/Helm (Enterprise) & Ground-control ops \\
EOEPCA+~\cite{eoepca} & -- & -- & CWL $\rightarrow$ Argo (Calrissian) & EO data pipelines \\
Kratos OpenSpace~\cite{kratos} & -- & -- & RKE2/K3s (ops) & Ground station infra \\
OPA Gatekeeper~\cite{gatekeeper} & -- & \checkmark (OPA/Rego) & -- & K8s admission (general) \\
\bottomrule
\end{tabular}
\end{table*}

\section{System Architecture}

\subsection{Pipeline Overview}
\label{sec:overview}

\begin{figure}[tb]
\centering
\begin{tikzpicture}[
  node distance=0.55cm,
  box/.style={rectangle, draw, thick, rounded corners, minimum width=3.0cm, minimum height=0.5cm, font=\footnotesize, align=center},
  arrow/.style={-{Stealth[length=2mm]}, thick}
]
  \node[box, fill=black!5] (input) {Mission Plan YAML};
  \node[box, fill=black!12, below=of input] (schema) {Stage 1: Schema Validation\\(Pydantic v2)};
  \node[box, fill=black!20, below=of schema] (policy) {Stage 2: Policy Evaluation\\(OPA/Rego)};
  \node[box, fill=black!12, below=of policy] (ir) {Stage 3: WorkflowIntent IR};
  \node[box, fill=black!5, below left=0.7cm and -0.5cm of ir] (argo) {Argo Workflow\\YAML};
  \node[box, fill=black!5, below right=0.7cm and -0.5cm of ir] (kueue) {Kueue Job\\YAML};

  \draw[arrow] (input) -- (schema);
  \draw[arrow] (schema) -- (policy);
  \draw[arrow] (policy) -- (ir);
  \draw[arrow] (ir.south) -- ++(0,-0.25) -| (argo.north);
  \draw[arrow] (ir.south) -- ++(0,-0.25) -| (kueue.north);
\end{tikzpicture}
\caption{Four-Stage Compilation Pipeline}
\label{fig:pipeline}
\end{figure}
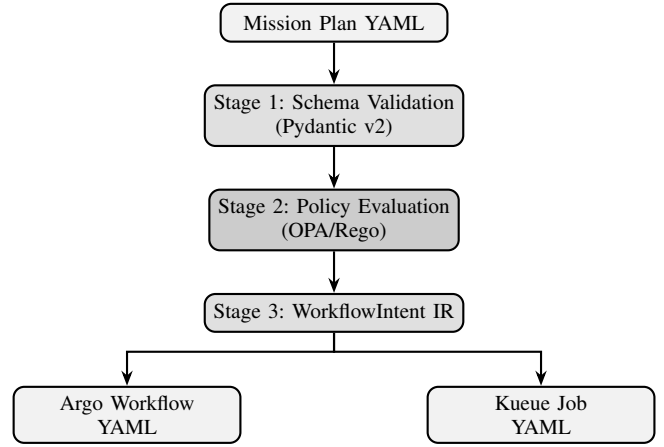

The compiler implements a four-stage pipeline (Fig.~\ref{fig:pipeline}): Stage~1 validates the mission plan schema using \texttt{Pydantic}, Stage~2 evaluates OPA/Rego policy rules, Stage~3 compiles validated plans into a typed \texttt{WorkflowIntent} intermediate representation (IR), and Stage~4 renders the IR into Argo Workflow and Kueue Job artifacts.

\subsection{Mission Plan Input Format}

Listing~\ref{lst:plan} shows an excerpt from a maritime surveillance mission plan. The plan declares two events on one orbit: an acquisition event at 10:30 running a three-step ship detection pipeline (preprocess, detect-ships, postprocess), and a download event at 10:42 transmitting results during a 268-second ground-visibility window. The \texttt{detect-ships} step requests GPU acceleration with CPU fallback.

\begin{figure}[t]
\begin{lstlisting}[caption={Mission plan input (abbreviated)},label={lst:plan},language=]
mission_id: mission-alpha
events:
  - timestamp: "2026-04-15T10:30:00Z"
    event_type: acquisition
    instrument: optical-camera
    services:
      - service_id: maritime-surveillance
        priority: 90
        steps:
          - name: preprocess
            image: ghcr.io/example/preprocess:0.1.0
            resource_class: cpu
          - name: detect-ships
            image: ghcr.io/example/ship-detector:0.1.0
            resource_class: gpu
            fallback_resource_class: cpu
          - name: postprocess
            resource_class: cpu
  - timestamp: "2026-04-15T10:42:00Z"
    event_type: download
    duration_seconds: 268
    ground_visibility: true
\end{lstlisting}
\end{figure}

\subsection{Schema Validation (Stage 1)}
\label{sec:schema}

The mission plan schema is implemented as \texttt{Pydantic}~v2~\cite{pydantic} \texttt{BaseModel} classes derived from publicly available ORCHIDE materials, with the structural elements summarized inline in Section~\ref{sec:orchide-background}. Key constraints enforced at parse time:

\begin{itemize}
  \item Acquisition events must specify an instrument.
  \item Download events must specify duration, must not declare AI services, and must require ground visibility.
  \item \texttt{mission\_id} must not be empty; \texttt{events} must contain at least one entry; each service must have at least one step.
\end{itemize}

Error messages include source references back to the ORCHIDE materials enumerated in Section~\ref{sec:orchide-background}, enabling downstream auditing of rule provenance. Service priority uses a 0--100 scale internally (0 is a misconfiguration, since ORCHIDE's target scale starts at~1, and is rejected by policy Rule~5); ORCHIDE's onboard scheduler uses 1--4 (Section~\ref{sec:orchide-background}). We preserve priority intent as-is in the domain model; translation to ORCHIDE's scale belongs in the rendering layer, which is a deployment-specific concern.

\subsection{Policy Evaluation (Stage 2)}
\label{sec:policy}

Ten OPA/Rego deny rules enforce semantic constraints, several requiring cross-field reasoning, as summarized in Table~\ref{tab:rules}.

\begin{table}[htbp]
\caption{OPA/Rego Policy Rules}
\label{tab:rules}
\centering
\small
\begin{tabular}{@{}clll@{}}
\toprule
\# & Constraint & Origin & Prov. \\
\midrule
1 & \texttt{mission\_id} not empty & Integrity & A \\
2 & At least one event & Integrity & A \\
3 & ACQ must declare services & ORCHIDE doc & D \\
4 & GPU/FPGA must have fallback & Safety & A \\
5 & Priority must not be zero & Range & A \\
6 & CPU must not claim accel. & Logic & A \\
7 & DL must not carry services & ORCHIDE doc & D \\
8 & DL requires visibility & ORCHIDE doc & D \\
9 & Services must have $\geq$1 step & Validity & A \\
10 & Landscape type recognized & ORCHIDE doc & D \\
\bottomrule
\multicolumn{4}{@{}l}{\scriptsize D = directly traceable to ORCHIDE docs; A = author-imposed safety policy}
\end{tabular}
\end{table}

Schema validation and policy evaluation intentionally overlap on Rules~1, 2, 7, 8, and~9 in the safety defense-in-depth sense of Section~\ref{sec:policy-rationale}: schema catches structural errors at parse time, while policy catches data bypassing schema validation (e.g., raw JSON submitted directly to OPA).

Listing~\ref{lst:rego} shows two representative Rego rules. Rule~4 requires any GPU or FPGA step (independent of the optional \texttt{needs\_acceleration} flag) to declare a fallback resource class. Rule~6 flags the contradictory combination of CPU resource class with \texttt{needs\_acceleration}.

\begin{figure}[t]
\begin{lstlisting}[caption={OPA/Rego policy rules (abbreviated)},label={lst:rego},language=]
accel_bound(s) if s.resource_class == "gpu"
accel_bound(s) if s.resource_class == "fpga"
missing_fallback(s) if not s.fallback_resource_class
missing_fallback(s) if s.fallback_resource_class == null

deny contains msg if {
  some i, j, k
  step := input.events[i].services[j].steps[k]
  accel_bound(step)
  missing_fallback(step)
  msg := sprintf("accelerator step %q "
    "(resource_class %q) must declare "
    "fallback_resource_class",
    [step.name, step.resource_class])
}

deny contains msg if {
  some i, j, k
  step := input.events[i].services[j].steps[k]
  step.resource_class == "cpu"
  step.needs_acceleration == true
  msg := sprintf("step %q claims needs_acceleration"
    " but uses cpu resource class", [step.name])
}
\end{lstlisting}
\end{figure}

OPA is invoked as a subprocess with a 30-second timeout (CWE-400 mitigation). Standard output and standard error are captured separately to prevent error message leakage into policy results.

\subsection{Compilation to IR (Stage 3)}

The compiler filters acquisition events and generates a \texttt{Work\-flow\-Intent} for each AI service. The IR carries ten resource hints including event timestamp, ground visibility, region type, orbit, duration, landscape type, execution mode, and three Boolean flags for GPU/FPGA requirements and fallback.

Workflow names are normalized to RFC~1123~\cite{rfc1123} DNS labels at IR creation time to ensure consistent resource identification across heterogeneous renderers.

\subsection{Argo Workflow Rendering (Stage 4a)}

The renderer produces Argo \texttt{v1alpha1} Workflow manifests with DAG-based execution. Sequential mode builds a linear dependency chain; parallel mode omits dependencies. Unknown execution modes default to sequential to prevent accidental parallelism.

Each step template includes phase annotations (preprocessing, AI inference, postprocessing) as defined in Section~\ref{sec:orchide-background}, and GPU steps use soft-preference node affinity with fallback via an environment variable. Both the raw 0--100 priority and its deterministic translation to ORCHIDE's 1--4 scale are emitted as workflow annotations (\texttt{orbital/priority} and \texttt{orbital/orchide-priority}). The translation buckets 0--100 onto 1--4 (76--100$\to$1 through 1--25$\to$4; priority~0 is rejected by Rule~5), taking 1 as the highest priority---the common scheduling convention, since ORCHIDE's public materials document the 1--4 scale but not its direction, so this too is an assumption pending interface disclosure. The \texttt{orbital/orchide-priority} key, and the assumption that an ORCHIDE Workflow Manager consumes it, are this paper's proposal pending disclosure of the deployment interface (Section~\ref{sec:orchide-background}); interoperation with the unreleased onboard scheduler is not verified here. Listing~\ref{lst:argo} shows a rendered DAG.

\begin{figure}[t]
\begin{lstlisting}[caption={Rendered Argo Workflow DAG (abbreviated)},label={lst:argo},language=]
apiVersion: argoproj.io/v1alpha1
kind: Workflow
metadata:
  name: mission-alpha-maritime-...-00z
  annotations:
    orbital/priority: '90'
    orbital/orchide-priority: '1'
    orbital/execution-mode: sequential
    orbital/requires-gpu: 'true'
spec:
  entrypoint: main
  templates:
  - name: main
    dag:
      tasks:
      - name: preprocess
        template: step-0-preprocess
      - name: detect-ships
        template: step-1-detect-ships
        depends: preprocess
      - name: postprocess
        template: step-2-postprocess
        depends: detect-ships
\end{lstlisting}
\end{figure}

\subsection{Kueue Job Rendering (Stage 4b)}

The renderer produces \texttt{batch/v1} Job manifests with Kueue~\cite{kueue} admission labels. Unlike the Argo renderer, which preserves the full multi-step DAG, the Kueue renderer generates an admission-oriented Job for the primary compute step (GPU step if present, else first step); full pipeline execution is delegated to the Argo path. Listing~\ref{lst:kueue} shows a rendered Job.

We chose Kueue over Volcano~\cite{volcano} for admission control. Kueue's ClusterQueue + LocalQueue abstraction handles fair queuing and GPU quotas without an additional scheduler. Kueue v0.17.0 introduced alpha-level DRA support for extended resources. Our renderer generates DRA ResourceClaims for GPU workloads and standard resource requests for CPU-only workloads.

\begin{figure}[t]
\begin{lstlisting}[caption={Rendered Kueue Job with DRA (abbreviated)},label={lst:kueue},language=]
apiVersion: batch/v1
kind: Job
metadata:
  generateName: mission-alpha-maritime-...-
  labels:
    kueue.x-k8s.io/queue-name: orbital-demo-local
    mission-id: mission-alpha
spec:
  template:
    spec:
      containers:
      - name: detect-ships
        resources:
          requests:
            cpu: '1'
            memory: 256Mi
          claims:
          - name: gpu
      resourceClaims:
      - name: gpu
        resourceClaimTemplateName: ...-gpu-claim
\end{lstlisting}
\end{figure}

\subsection{Loss Events and Severity Tiers}
\label{sec:loss-events}

A ground-side admission-gate compiler is only useful insofar as the checks it enforces correspond to real in-orbit consequences. We therefore classify the loss events that a malformed or under-constrained mission plan can produce into four severity tiers, each tied to the schema or policy check that catches it. The classification is consequence-oriented, in the spirit of hazard-tier practice in software-safety engineering---the severity classification of NASA-STD-8739.8B~\cite{nasa-std-8739-8} and the constraint-enforcement perspective of STAMP~\cite{leveson-stamp}, whose associated hazard-analysis technique is STPA---but the tiers below are an operational classification for the compiler's admission gate, not a flight safety case. Provenance tags in Table~\ref{tab:loss-events} reproduce the convention of Table~\ref{tab:rules}: \textbf{D} denotes a check derived from ORCHIDE's published Deferred-Phase scope, and \textbf{A} denotes a check that the authors impose as safety practice.

\begin{table*}[!t]
\caption{Pre-Uplink Loss-Event Severity Tiers and Their Schema/OPA Checks}
\label{tab:loss-events}
\centering
\small
\begin{tabular}{@{}cllcl@{}}
\toprule
Tier & Loss event & Detected by & Rules & In-orbit consequence \\
\midrule
T1 & Plan malformed                & Pydantic + OPA    & 1, 2      & none (ground reject) \\
T2 & Accelerator stall, deferrable & OPA               & 4, 6      & Deferred-Phase stall \\
T3 & Contact-window loss           & Pydantic + OPA    & 7, 8      & missed/malformed downlink \\
T4 & Mission-objective loss        & OPA               & 3, 5, 9, 10 & AI service drop \\
\bottomrule
\end{tabular}

\smallskip
\noindent\scriptsize Provenance per row: T1\,\textbf{A} (Rules 1\,\textbf{A}, 2\,\textbf{A}); T2\,\textbf{A} (Rules 4\,\textbf{A}, 6\,\textbf{A}); T3\,\textbf{D} (Rules 7\,\textbf{D}, 8\,\textbf{D}); T4 mixed (Rule 3\,\textbf{D}, Rule 5\,\textbf{A}, Rule 9\,\textbf{A}, Rule 10\,\textbf{D}). Rule~5's \textbf{A} tag reflects that its priority$\,\neq\,$0 check operates on the compiler's 0--100 scale rather than ORCHIDE's documented 1--4 scale; its Tier~T4 placement is interpretive. The \emph{Detected by} column lists the layer(s) that catch every rule in the tier; in T4, Rule~9 is also schema-enforced (the Section~\ref{sec:policy-rationale} overlap set), but Rules~3, 5, and~10 are policy-only, so the tier as a whole is caught only by OPA.
\end{table*}

\paragraph*{Tier interpretation and provenance.} T1 captures structural violations: missing required fields, malformed timestamps, or type errors caught by the \texttt{Pydantic} schema, plus the OPA backstops for an empty \texttt{mission\_id} (Rule~1) and a plan with no events (Rule~2). Such plans are rejected on the ground before any uplink is attempted; the in-orbit consequence is therefore null. T1 is author-imposed (\textbf{A}); Rules~1--2 narrow the residual risk on a schema-bypass path, but if both Pydantic and the policy package are bypassed entirely no pre-uplink guarantee applies. T2 covers semantic violations that would not crash the onboard Orchestrator but could cause the Deferred Phase to stall: a GPU step with no \texttt{fallback\_resource\_class} (Rule~4) cannot be rescheduled if the accelerator is unavailable, and a CPU step asserting \texttt{needs\_acceleration} (Rule~6) is internally contradictory. Both are author-imposed (\textbf{A}); ORCHIDE documents the field but does not prescribe how the Orchestrator should act on it (Section~\ref{sec:orchide-background}). ``Deferred-Phase stall'' is one plausible behavior under that unspecified policy; the rule prevents the ambiguous configuration from reaching orbit regardless of the Orchestrator's eventual behavior. T3 covers checks tied to the download-event contract that ORCHIDE's published materials specify directly (\textbf{D}): a download event carrying AI services (Rule~7) violates the published ORCHIDE invariant that \textsc{download} events carry no \textsc{workflow} field (Section~\ref{sec:orchide-background}), and a download event with missing or false \texttt{ground\_visibility} (Rule~8) cannot transmit during its scheduled window. T4 captures checks whose violation can drop AI processing of a primary observation. Rule~3 is per-event: an \emph{acquisition} event with no services attached (\textbf{D}) is still acquired at the sensor but produces no processed output. Rules~5, 9, and~10 are per-service (Rego \texttt{some i, j}): priority zero on the compiler's 0--100 scale, a range check informed by ORCHIDE's documented 1--4 scale (Rule~5, \textbf{A}), a service with no executable steps (Rule~9, \textbf{A}), or a service with an unrecognized \texttt{landscape\_type} preventing model selection (Rule~10, \textbf{D}; the field is optional, so the rule flags a present unrecognized value but permits a null or omitted one) each drop the affected service. None of these consequence modes involves spacecraft hardware damage---the failure modes are scheduling stalls, missed contact windows, or null AI-service execution. These in-orbit consequences are derived from ORCHIDE's published Deferred-Phase semantics rather than measured on the onboard runtime, which is not publicly available; the tiers are a mechanistically grounded pre-uplink risk model that motivates the admission gate, not empirically measured outcomes.

\section{MCP Agent Interface}

Six Model Context Protocol~\cite{mcp-spec} tools expose the pipeline to AI agents via FastMCP~\cite{fastmcp}: \texttt{validate\_plan} (schema check), \texttt{compile\_plan} (IR generation), \texttt{render\_argo} (workflow rendering), \texttt{explain\_policy} (OPA evaluation), \texttt{diff\_plans} (structural comparison), and \texttt{check\_timeline\_conflicts}. The last tool detects interval overlaps using $\max(a_{\mathrm{start}}, b_{\mathrm{start}}) < \min(a_{\mathrm{end}}, b_{\mathrm{end}})$. A scripted client exercises the tools end-to-end through the real MCP server in \texttt{scripts/mcp\_agent\_demo.py}: it validates a plan, surfaces a missing-fallback violation via \texttt{explain\_policy} (Rule~4), then re-checks a corrected plan and compiles and renders it. The fix is applied from a pre-authored corrected plan rather than synthesized by an LLM; the demonstration shows the tool surface is composable and agent-consumable, and closing the diagnose--repair--verify loop with an autonomous agent is future work.

File path validation uses \texttt{Path.relative\_to()} containment checks rather than prefix matching to prevent path traversal (CWE-22). Plan files accept only bare filenames within a whitelisted directory.

\section{Evaluation}

\subsection{Case Study: Error Detection}
\label{sec:case-study}

We evaluated the compiler on three categories of mission plans. Table~\ref{tab:casestudy} summarizes the results.

\begin{table*}[htbp]
\caption{Case Study: Error Detection Across Plan Categories}
\label{tab:casestudy}
\centering
\small
\begin{tabular}{@{}llccp{4.8cm}@{}}
\toprule
Category & Example Error & Sch. & Pol. & Notes \\
\midrule
Valid (ORCHIDE, 3 events) & None & \checkmark & \checkmark & 3 Argo WFs + 3 Kueue Jobs \\
\midrule
Struct: DL w/o visibility & \texttt{visibility=false} & \checkmark & \checkmark & Both (Rule 8) \\
Struct: ACQ w/o instrument & Missing \texttt{instrument} & \checkmark & -- & Schema-only \\
Struct: empty \texttt{mission\_id} & \texttt{mission\_id=""} & \checkmark & \checkmark & Both (Rule 1) \\
Struct: DL with services & Services on DL event & \checkmark & \checkmark & Both layers catch \\
\midrule
Semantic: invalid \texttt{landscape\_type} & \texttt{landscape\_type=desert} & -- & \checkmark & Rule 10: unknown value \\
Semantic: GPU no fallback & No \texttt{fallback\_class} & -- & \checkmark & Rule 4: safety \\
Semantic: CPU+accel & \texttt{cpu}+\texttt{accel} & -- & \checkmark & Rule 6: within-step \\
\bottomrule
\end{tabular}
\end{table*}

\textbf{Valid~plans:} The ORCHIDE-format sample plan (3 events, 3 AI services across 2 orbits) passes both schema and policy validation and compiles to 3 Argo Workflows and 3 Kueue Jobs.

\textbf{Structural~errors:} Four categories of structurally invalid plans were rejected at schema validation (Stage~1). Three of the four are independently caught by policy as well---empty \texttt{mission\_id} (Rule~1), download without visibility (Rule~8), and download with services (Rule~7)---demonstrating the layered schema--policy coverage of Section~\ref{sec:policy-rationale}; only acquisition without instrument is schema-exclusive.

\textbf{Semantic~errors:} Three categories of semantic violations passed schema validation but were flagged by OPA policy (Rules~4, 6, 10). A Pydantic \texttt{model\_validator} could express these checks (the schema already uses one for the download-event constraints, Table~\ref{tab:casestudy}), so we place them in the Rego policy not for expressiveness but for governance: the policy can be executed and audited standalone by a reviewer who never runs the compiler (Section~\ref{sec:policy-rationale}).

\subsection{Compilation Performance and Scaling}

Table~\ref{tab:perf} shows phase-wise compilation times for synthetic mission plans with 10 to 1000 acquisition events (3 steps each), measured on an Intel Core i5-7400 @ 3.00\,GHz (single core, Ubuntu 24.04, Python 3.12.3, OPA 1.15.1), mean of 30 iterations.

\begin{table}[htbp]
\caption{Phase-Wise Compilation Time (ms, mean $\pm$ std)}
\label{tab:perf}
\centering
\scriptsize
\begin{tabular}{@{}rrrrrr@{}}
\toprule
Events & Parse & OPA & Compile & Render & Total \\
\midrule
10 & 14.1$\pm$3.9 & 18.8$\pm$3.2 & 0.2$\pm$0.0 & 0.5$\pm$0.3 & 33.7$\pm$6.4 \\
50 & 64.6$\pm$10.8 & 21.5$\pm$3.9 & 0.9$\pm$0.1 & 2.4$\pm$1.3 & 89.4$\pm$13.7 \\
100 & 127$\pm$7 & 27.7$\pm$4.0 & 1.8$\pm$0.4 & 5.2$\pm$2.0 & 162$\pm$8 \\
500 & 641$\pm$16 & 68.4$\pm$6.1 & 11.2$\pm$6.5 & 28.4$\pm$7.7 & 749$\pm$19 \\
1000 & 1323$\pm$36 & 115$\pm$5 & 20.9$\pm$9.2 & 62.7$\pm$14.5 & 1522$\pm$44 \\
\bottomrule
\end{tabular}
\end{table}

Parse and total time scale near-linearly; parse (YAML + Pydantic) dominates at large sizes (its share grows from 41.8\% at 10 events to 86.9\% at 1000) while the OPA subprocess scales sub-linearly (18.8\,ms $\rightarrow$ 115\,ms), its per-call cost being a large fixed component (process spawn and policy load) plus a smaller input-proportional term. The sub-100\,ms compile and render phases carry large relative variance because this host also runs a single-node kubeadm cluster; background control-plane load on this shared host, rather than the compiler, likely drives that jitter, so we read them as order-of-magnitude costs. A 1000-event plan with 3000 steps compiles in about 1.5\,seconds, dominated by the pure-Python YAML load. Benchmarking OPA against an in-process Python validator (\texttt{baseline\_validator.py}) that re-implements the ten rules and reproduces OPA's decision on the entire ablation corpus (fourteen cases spanning valid, schema-only, policy-only, and both-layer inputs), the latter runs in 0.02--1.0\,ms versus 17--120\,ms in the same paired run. This run is separate from Table~\ref{tab:perf}, so its OPA figures differ from that table's 18.8--115\,ms through ordinary run-to-run variance on the shared host. The gap over the in-process baseline (870$\times$ at 10 events, 120$\times$ at 1000) is largest where that fixed per-call overhead dominates; a long-lived OPA server or embedded engine would remove the fixed component while preserving standalone auditability. We keep the subprocess model because $\sim$120\,ms is negligible beside parse; the baseline is a performance reference and equivalence oracle, not a rejected design.

\subsection{Defense-in-Depth Analysis}

Whereas Table~\ref{tab:casestudy} in Section~\ref{sec:case-study} shows specific example errors for each constraint category, Table~\ref{tab:overlap} below summarizes, for eight representative constraint categories, which layer catches each independently of the example workload. Three of the eight are enforced by both layers; the structural Rules~1 and~2 (\texttt{mission\_id}, event count) are two further schema--policy overlaps not tabulated in this event/service/step-level subset. If the schema layer were removed, seven of eight constraint categories would still be caught by policy. If the policy layer were removed, four semantic constraints (GPU fallback, CPU acceleration contradiction, priority validation, and landscape type) would go unchecked.

\begin{table}[htbp]
\caption{Defense-in-Depth: Schema vs. Policy Overlap}
\label{tab:overlap}
\centering
\small
\begin{tabular}{@{}l|c|c|c@{}}
\toprule
Constraint & Sch. & Pol. & Both \\
\midrule
ACQ needs instrument & \checkmark & -- & -- \\
DL no services & \checkmark & \checkmark & \checkmark \\
DL needs visibility & \checkmark & \checkmark & \checkmark \\
Service $\geq$1 step & \checkmark & \checkmark & \checkmark \\
GPU/FPGA needs fallback & -- & \checkmark & -- \\
CPU contradicts accel & -- & \checkmark & -- \\
Priority $\neq$ 0 & -- & \checkmark & -- \\
Landscape recognized & -- & \checkmark & -- \\
\bottomrule
\end{tabular}
\end{table}

\subsection{Live Cluster Validation}
\label{sec:live-cluster}

We validated rendered artifacts on a live single-node K8s~v1.35.1 cluster (8~CPU, 32\,GiB, NVIDIA RTX~5080 GPU) running Argo~v4.0.1 and Kueue~v0.17.0. Four experiments are reported in this subsection; a fifth experiment, exercising the explicit DRA Template Path on the same host after a platform upgrade, is presented in Section~\ref{sec:dra-quota}.

\noindent\textbf{Argo~sequential~DAG:} A three-step CPU-only mission plan was compiled and submitted. All steps executed in correct DAG order, completing with Succeeded~3/3.

\textbf{Real~GPU~allocation:} A mission plan with a CUDA container (\texttt{nvidia/\allowbreak cuda:13.0.0-base-ubuntu22.04}) was compiled and submitted using DRA via the \texttt{gpu.nvidia.com} DeviceClass. The GPU step executed \texttt{nvidia-smi}, confirming RTX~5080 allocation (16,303\,MiB, CUDA~13.0, Driver~580.126). Workflow completed Succeeded~3/3.

\textbf{GPU~path~comparison:} The same CUDA container was submitted twice: once with a DRA ResourceClaim (output: \texttt{PATH=GPU}, RTX~5080 detected) and once without (output: \texttt{PATH=CPU-FALLBACK}, no GPU device). This shows that DRA claims control GPU device visibility and that the runtime fallback path works end-to-end.

\textbf{Kueue~quota~contention:} Two CPU Jobs (priority~90 and~50) were submitted simultaneously to a ClusterQueue with CPU quota~4. Each requested~3~CPUs (combined 6$>$4), so Kueue admitted the first and suspended the second. Upon completion, the second was automatically admitted within 1\,second. Both completed successfully.

\subsection{DRA-Backed GPU Admission on a Live Cluster}
\label{sec:dra-quota}

To demonstrate that Kueue-rendered Job artifacts compose correctly with the DRA path on a recent stable Kueue release, we ran a quota-and-admission experiment that exercises both layers (DRA device binding by the scheduler and ClusterQueue quota accounting by Kueue) on a single-device GPU queue.

\noindent\textbf{Setup:} The experiment ran on a single-node host kubeadm Kubernetes v1.36.1 cluster with one NVIDIA GeForce GT~1030 device, Cilium CNI, Multus secondary CNI, and the NVIDIA DRA driver~\cite{nvidia-dra-driver}, which NVIDIA donated to the Cloud Native Computing Foundation at KubeCon~EU~2026~\cite{nvidia-dra-cncf}; the GT~1030 is used purely as a DRA-allocatable device whose role is to exercise Kueue's quota-and-binding cycle (no GPU-compute workload is run against it). The same physical host previously ran the experiments in Section~\ref{sec:live-cluster} at Kubernetes v1.35.1 with an RTX~5080; reproducing the experiment in this subsection requires only Kubernetes v1.36.1 and a single DRA-allocatable GPU. Kueue~v0.17.3~\cite{kueue,kueue-v0173} ran with the \texttt{DRA\allowbreak Extended\allowbreak Resources} and \texttt{Dynamic\allowbreak Resource\allowbreak Allocation} feature gates enabled on the controller-manager and a Configuration ConfigMap whose \texttt{device\allowbreak Class\allowbreak Mappings} entry bound the DRA DeviceClass \texttt{gpu.\allowbreak nvidia.\allowbreak com} to a logical resource name \texttt{dra.\allowbreak gpu.\allowbreak nvidia.\allowbreak com}~\cite{kueue-dra-extended-resources}. A ClusterQueue was provisioned covering \texttt{cpu}, \texttt{memory}, and \texttt{dra.\allowbreak gpu.\allowbreak nvidia.\allowbreak com}, with \texttt{nominalQuota} of one GPU.

\noindent\textbf{Methodology:} Two Kubernetes Jobs were submitted simultaneously to the LocalQueue backing the ClusterQueue. Each Pod declared a \texttt{resourceClaims} entry pointing to a single-GPU \texttt{Resource\allowbreak Claim\allowbreak Template} and a container that referenced the claim via \texttt{resources.claims}; the template requested one device from \texttt{deviceClassName:\allowbreak gpu.\allowbreak nvidia.\allowbreak com}. We checked four properties: (i)~at most one non-Finished Workload is in \texttt{Admitted=True} at any time (the \texttt{Admitted} status is monotone and persists after \texttt{Finished=True}; quota release is bound to \texttt{Finished}, not to clearing \texttt{Admitted}); (ii)~the second Workload remains pending until quota is released; (iii)~Kueue accounts the allocation against the \texttt{device\allowbreak Class\allowbreak Mappings} logical resource so \texttt{flavorsUsage[\allowbreak dra.\allowbreak gpu.\allowbreak nvidia.\allowbreak com]=1}; (iv)~upon completion of the first Job, the queued Workload transitions to \texttt{Admitted=True} without operator intervention, the DRA driver instantiates a fresh \texttt{Resource\allowbreak Claim} (\texttt{state=allocated, reserved}), and the second Job runs to completion.

\noindent\textbf{Results:} Both Jobs were accepted by the API server. At submission both Workloads were Suspended (\texttt{pending=2,\allowbreak admitted=0}); within $\sim$10 seconds Kueue admitted one (\texttt{pending=1,\allowbreak admitted=1,\allowbreak flavorsUsage[\allowbreak dra.\allowbreak gpu.\allowbreak nvidia.\allowbreak com]=1}). The first Job logged \texttt{job-1 starting at 18:04:15} and \texttt{job-1 done at 18:04:45}; the queued Workload then transitioned to \texttt{Admitted=True} after a $\sim$5-second cascade gap (wall-clock between job-1 container exit and job-2 container start; not decomposed by stage), with a freshly created \texttt{Resource\allowbreak Claim} bound to the GT~1030. The second Job logged \texttt{job-2 starting at 18:04:50} and \texttt{job-2 done at 18:05:20}. Final state: both Workloads \texttt{Admitted=True,\allowbreak Finished=True}; both Jobs \texttt{Complete 1/1}. All four properties held without manual reconciliation.

\noindent\textbf{Scope of this result:} What was validated is the explicit DRA Template Path: a per-Pod \texttt{Resource\allowbreak Claim} created from a \texttt{Resource\allowbreak Claim\allowbreak Template}, its quota accounted via \texttt{device\allowbreak Class\allowbreak Mappings}, and the admission cascade triggered on first-Job completion. The experiment does not measure scheduling fairness beyond two contestants and does not exercise multi-class device flavors (e.g., mixed FPGA+GPU). The setup is for Kueue~v0.17.3; in v0.18 the corresponding gates were renamed to \texttt{Kueue\allowbreak DRA\allowbreak Integration} (Beta, default-on) and \texttt{Kueue\allowbreak DRA\allowbreak Integration\allowbreak Extended\allowbreak Resource} (Alpha, default-off), with the legacy \texttt{Dynamic\allowbreak Resource\allowbreak Allocation}/\allowbreak\texttt{DRA\allowbreak Extended\allowbreak Resources} names deprecated and locked-to-default in v0.18 and scheduled for removal in v0.19~\cite{kueue-dra-extended-resources}. Section~\ref{sec:dra-unified} re-validates the renderer against the renamed \texttt{Kueue\allowbreak DRA\allowbreak Integration} gate on v0.18.3 and extends this single-class result to a unified GPU+CPU quota with a scheduler-level accelerator fallback.

\noindent\textbf{Reproducibility:} The Configuration patch, ClusterQueue+\allowbreak LocalQueue+\allowbreak ResourceFlavor, \texttt{Resource\allowbreak Claim\allowbreak Template}, and the two-Job manifest are committed at \texttt{manifests/\allowbreak k8s/\allowbreak kueue/\allowbreak dra-paper-test/} as files 00 through 03, with the captured experimental output in \texttt{docs/\allowbreak experiments/\allowbreak 2026-06-09-\allowbreak dra-quota-\allowbreak cascade-output.md}. The deterministic CPU pipeline checks (schema, policy, render, Argo submit and completion, Kueue Job render and admission, Kueue Job completion) are exercised by \texttt{scripts/\allowbreak validate\_live\_cluster.sh} on the CPU mission plan \texttt{configs/\allowbreak mission\_plans/\allowbreak validation\_live\_cluster.yaml}, returning \texttt{PASS:~13 FAIL:~0}.

\subsection{Unified GPU+CPU DRA and Scheduler-Level Fallback}
\label{sec:dra-unified}

The Section~\ref{sec:dra-quota} experiment counted a single GPU device class under one ClusterQueue. This subsection extends that result in three directions on the same host after upgrading Kueue to v0.18.3~\cite{kueue-v0183} (Kubernetes~v1.36.1, one NVIDIA Quadro~K2200 in the single-DRA-allocatable-GPU role of Section~\ref{sec:dra-quota}, and \texttt{kubernetes-sigs/\allowbreak dra-driver-cpu}~\cite{dra-cpu-driver} publishing a \texttt{dra.\allowbreak cpu} DeviceClass): (i)~a \emph{unified} quota that counts a CPU device class alongside the GPU class; (ii)~a \emph{scheduler-level} accelerator fallback that replaces the runtime environment-variable switch of Section~\ref{sec:live-cluster}; and (iii)~the empirical boundary between them. On v0.18.3 the DRA integration is governed by the \texttt{Kueue\allowbreak DRA\allowbreak Integration} gate (Beta, default-on), so no explicit feature-gate flag is required; this re-validates the renderer against the renamed gates that Section~\ref{sec:dra-quota} left open.

\noindent\textbf{Unified quota (both classes counted).} We extended the Configuration \texttt{device\allowbreak Class\allowbreak Mappings} to bind \texttt{gpu.\allowbreak nvidia.\allowbreak com} to \texttt{dra.\allowbreak gpu.\allowbreak nvidia.\allowbreak com} \emph{and} \texttt{dra.\allowbreak cpu} to a second logical resource \texttt{dra.\allowbreak cpu}, and provisioned a ClusterQueue covering both plus \texttt{cpu} and \texttt{memory}. Two Jobs each claiming one \texttt{dra.\allowbreak cpu} device through an \texttt{exactly} request reproduced the Section~\ref{sec:dra-quota} cascade for the CPU class: Kueue admitted the first (\texttt{flavorsUsage[\allowbreak dra.\allowbreak cpu]=1}) and gated the second with the explicit reason ``insufficient unused quota for \texttt{dra.\allowbreak cpu} \dots\ 1 more needed,'' then cascade-admitted it on completion. The GPU cascade of Section~\ref{sec:dra-quota} reproduced identically on v0.18.3 (\texttt{flavorsUsage[\allowbreak dra.\allowbreak gpu.\allowbreak nvidia.\allowbreak com]=1}). Each accelerator class thus carries its own logical quota under a single ClusterQueue, subject to the boundary below.

\noindent\textbf{Scheduler-level fallback (\texttt{firstAvailable}).} Kubernetes DRA expresses ``prefer GPU, else CPU'' declaratively through a \texttt{firstAvailable} ResourceClaim request---a prioritized subrequest list that graduated to stable in Kubernetes~v1.36 (KEP-4816~\cite{dra-prioritized-list}). This is a scheduler-level decision, in contrast to the Section~\ref{sec:live-cluster} runtime switch in which the container inspects an environment variable at start-up. Two identical plain Pods, each referencing one \texttt{firstAvailable[\allowbreak gpu.\allowbreak nvidia.\allowbreak com,\allowbreak dra.\allowbreak cpu]} claim, demonstrated the fallback: the first bound the K2200 (\texttt{device=gpu-0}, driver \texttt{gpu.\allowbreak nvidia.\allowbreak com}) and the second, finding the GPU taken, fell back to a CPU device (\texttt{device=cpudevnuma000}, driver \texttt{dra.\allowbreak cpu})---same specification, the scheduler choosing. The compiler emits this claim under an opt-in \texttt{-{}-dra-\allowbreak fallback} flag for a step whose primary and fallback resource classes are both driver-backed; FPGA is deliberately excluded, as no FPGA DRA driver exists. Upstream KEP-4816 examples prioritize tiers \emph{within} one GPU driver; we did not find an upstream example demonstrating this cross-driver GPU$\rightarrow$CPU form, so it is an application the specification permits rather than one it illustrates.

\noindent\textbf{The boundary (\texttt{firstAvailable} is not admissible under Kueue).} Kueue quota counting supports only \texttt{exactly} device requests~\cite{kueue-dra-extended-resources}. Submitting the same \texttt{firstAvailable} claim \emph{as a Kueue Job} makes the boundary concrete: Kueue rejects the Workload as Inadmissible (\texttt{Quota\allowbreak Reserved=\allowbreak False, reason=\allowbreak Inadmissible}) with the message ``\texttt{Resource\allowbreak Claim\allowbreak Template} \dots: FirstAvailable device selection is not supported''; the Job stays Suspended and nothing is quota-counted. We observed the identical rejection on both Kueue~v0.17.3 and v0.18.3, so it is a property of the integration rather than of one release. The two mechanisms are therefore disjoint under current Kueue: the scheduler-level \texttt{firstAvailable} fallback lives off the admission path (a plain Pod or the Argo route), while a Kueue-admitted Job uses an \texttt{exactly} claim. The renderer reflects this boundary: for a Kueue Job it emits an \texttt{exactly} \texttt{gpu.\allowbreak nvidia.\allowbreak com} claim (which Kueue quota-counts) and never a \texttt{firstAvailable} claim; applying the renderer's \texttt{-{}-dra-\allowbreak fallback} output to the live queue yielded an admitted, quota-counted Workload (\texttt{Admitted=True}), confirming the path end-to-end.

\noindent\textbf{Reproducibility.} The Configuration patch, ClusterQueue, both \texttt{Resource\allowbreak Claim\allowbreak Template}s, and the three demonstrations (unified CPU cascade, \texttt{firstAvailable} fallback Pods, and \texttt{firstAvailable}-under-Kueue rejection) are committed at \texttt{manifests/\allowbreak k8s/\allowbreak kueue/\allowbreak dra-unified/}, with captured output under \texttt{results-\allowbreak v0.18.3-\allowbreak 20260716/}. The \texttt{dra-driver-cpu} install carries a documented kubelet-root-dir path fix---a local workaround for a Helm-chart limitation that remains open upstream.

\subsection{Test Suite and Static Validation}

The test suite spans every stage of the pipeline (schema validation, policy evaluation, compilation, rendering both sequential and parallel, the command-line interface, MCP tools and security, interface contracts, continuous integration, and Docker) and exercises all ten policy rules directly: the in-process baseline reproduces OPA's accept/reject decision across the entire ablation corpus, so every rule is checked at the decision level rather than only by line execution. In a fresh environment lacking OPA or a live cluster, the OPA- and cluster-dependent cases skip and the remainder pass; the full suite passes with the local OPA binary and a live single-node kubeadm cluster. Continuous integration runs the suite as a merge gate.

Golden evaluation fixtures verify end-to-end compilation correctness by comparing actual \texttt{WorkflowIntent} output against expected JSON. Rendered Argo Workflow YAML is validated by \texttt{argo lint}~\cite{argo} locally; CI runs Python-based manifest structure checks when the Argo CLI is unavailable.

\subsection{Security}

MCP path traversal protection is tested with six attack vectors including absolute paths, \texttt{..} traversal, directory components, and prefix bypass. OPA subprocess timeout is tested for CWE-400 mitigation.

\section{Limitations and Future Work}
\label{sec:limitations}

FPGA, mixed-accelerator, and multi-class DRA policies remain future work; DRA-backed GPU admission with single-device quota is documented in Section~\ref{sec:dra-quota}. All test mission plans were author-constructed; validation against real satellite operator mission plans would strengthen external validity. The system does not implement radiation hardening, full high availability (HA), or over-the-air (OTA) update guarantees, and does not implement security controls in the NIST~SP~800-53 sense (Section~\ref{sec:policy-rationale}); it complements onboard platforms rather than replacing them. Because \texttt{landscape\_type} is optional (Section~\ref{sec:loss-events}, Rule~10), a service that omits it while nonetheless depending on landscape context is admitted by both layers---the omission is indistinguishable at compile time from ``not applicable,'' and catching it would require runtime context the ground-side compiler lacks. ORCHIDE's archival paper~\cite{weisz2026orchide} identifies its own ground-side SDK as ongoing work.

\paragraph*{Accelerator fallback and Kueue admission.} The scheduler-level \texttt{firstAvailable} fallback and Kueue quota accounting are disjoint under current Kueue (Section~\ref{sec:dra-unified}): Kueue rejects a \texttt{firstAvailable} claim as Inadmissible and quota-counts only \texttt{exactly} requests, so expressing accelerator fallback and Kueue quota within a single claim awaits upstream Kueue support. The CPU DRA path depends on \texttt{kubernetes-sigs/\allowbreak dra-driver-cpu}~\cite{dra-cpu-driver}, which is pre-beta, and on a local kubelet-root-dir workaround for a Helm-chart limitation still open upstream; these are integration constraints of the current driver ecosystem rather than properties of the compiler. The cross-driver GPU$\rightarrow$CPU \texttt{firstAvailable} form, while permitted by KEP-4816, is not illustrated by upstream examples, so its portability across DRA drivers remains to be established.

\paragraph*{Generalization beyond ORCHIDE.} The framework treats ORCHIDE as the inaugural case study, not as a hard coupling. The \texttt{Pydantic} schema, the Rego policy package, and the Argo/Kueue renderer are independent modules; another operator's plan format can replace the schema, that operator's flight rules can replace or augment the policy package, and an alternative onboard runtime (for example, MicroShift~\cite{microshift-axdcu1} Jobs on AxDCU-1-class hardware, or raw CWL via Calrissian~\cite{eoepca} for an EOEPCA-aligned deployment) can replace the renderer without disturbing the compilation pipeline. The author-imposed safety rules (rows marked `A' in Table~\ref{tab:rules}) are designed to carry over directly (priority bounds, fallback-class requirements, and CPU/accelerator-coherence checks are not ORCHIDE-specific), though no second renderer is yet demonstrated, so this modularity is a design property, not an evaluated result. Unlike admission-time validators (OPA Gatekeeper~\cite{gatekeeper}, or Kubernetes-native CEL admission policies) that check rendered Kubernetes objects at API-server admission, this compiler checks mission-plan domain semantics before any Kubernetes object exists; the two are complementary.

\section{Conclusion}

We presented an open-source ground-side compiler that addresses the mission plan validation and compilation gap identified in onboard satellite platforms like ORCHIDE. The system applies OPA/Rego policy-as-code and Kueue-compatible rendering to satellite mission plans, producing Argo Workflow and Kueue Job artifacts verified through static lint checks and live cluster submission. A case study demonstrated that the layered validation approach catches both structural errors (at schema parse time) and semantic violations (via OPA policy), with end-to-end compilation scaling near-linearly (parse-dominated at scale) from 34\,ms (10 events) to 1.5\,s (1000 events). The tool is available under EUPL-1.2 at {\small\url{https://github.com/thc1006/satellite-mission-compiler}} (tag~v0.4.2, DOI~10.5281/\allowbreak zenodo.21228150).

\section*{Acknowledgment}

This work was supported in part by the Center for Intelligent Team Robotics and Human-Robot Collaboration under the ``Top Research Centers in Taiwan Key Fields Program'' of the Ministry of Education (MOE), Taiwan, and in part by the National Science and Technology Council, Taiwan, under grants 115-2221-E-A49-004 and 115-2640-E-011-003.

\bibliographystyle{IEEEtran}

\end{document}